\documentclass[aps,prl,floats,floatfix,twocolumn,amsfonts,nofootinbib]{revtex4}
\usepackage{bm}
\usepackage{graphicx}
\def\be{\begin{equation}}
\def\ee{\end{equation}}
\def\rb{\rangle}
\def\lb{\langle}
\def\Mf{~MeV-fm$^3$}
\begin{document}
\title{Spin-triplet pairing in large nuclei}

\author{G.F. Bertsch and Y. Luo}
\affiliation{Institute for Nuclear Theory and Dept. of Physics,
University of Washington, Seattle, Washington}

\date{12/11/2009}

\begin{abstract}

The nuclear pairing condensate is expected to change character from
spin-singlet to spin-triplet when the nucleus is very large and 
the neutron and proton numbers $Z,N$ are
equal. We investigate the transition between these
two phases within the framework of the Hartree-Fock-Bogoliubov equations,
using a zero-range interaction to generate the pairing.  We confirm
that extremely large nucleus would indeed favor triplet pairing 
condensates, with the Hamiltonian parameters taken from known 
systematics.  The favored 
phase is found to depend on the specific orbitals at the Fermi energy.  
The smallest nuclei with a well-developed spin-triplet condensate are in 
the mass region $A\sim 130-140$.\\
\end{abstract}

\maketitle

\section{Introduction}

The neutron-proton interaction in the spin-one channel 
is attractive and stronger than the identical-nucleon
interaction.   Nevertheless, the nuclei observed in nature favor
identical-particle pairing.  A trivial answer to this puzzle is
that most nuclei have different numbers of protons and neutrons,
and the isospin polarization discourages pairing in the isospin-zero
(T=0) channel.  But even in nuclei with equal numbers of neutron and
protons the $J=1,T=0$ neutron-proton pairing does not compete well
with ordinary $J=0,T=1$ pairing, as evidenced by the ground-state
spins and binding systematics of odd-odd nuclei\cite{ma00a,ma00b}.  
This puzzle presents a serious challenge to nuclear theory that must
be resolved if the theory is to be used confidently in contexts where
experimental information is not available, such as astrophysical
environments.  

It has recently been suggested that the explanation of the suppressed
spin-triplet pairing is the presence of the 
strong nuclear spin-orbit field
\cite{po98,baroni,schwenk,superconductivity}.  The spin-orbit field
interferes with the pairing in both channels, but it suppresses it more strongly
in the $J=1$ neutron-proton 
channel than in the $J=0$ identical-particle channel.  Since
the spin-orbit splitting is a surface effect, one might expect 
that the pairing would change character to the $J=1$ condensate in 
extremely large nuclei.  In the limit of large nuclear size, the
surface-to-volume ratio would be small and the spin-orbit field would
be ineffective at controlling the single-particle spectrum.
In this paper we apply the Hartree-Fock-Bogoliubov (HFB) theory with
a simplified Hamiltonian to examine the properties of the pairing condensate as one
approach this limit. We shall find that 
the transition to $J=1$ pairing does occur in large nuclei, but is only
strong for nuclei
that are somewhat beyond the limit of stability with respect to proton emission.

\section{The Hamiltonian}
To keep the calculations as simple as possible, we consider a
Hamiltonian consisting of a phenomenological single-particle Hamiltonian together
with a two-particle interaction,
\be
\hat H = \sum_i \lb i| H_{sp}| j \rb a^\dagger_i a_j +
\sum_{i>j,k>l}  \lb ij|v | kl\rb a^\dagger_i a^\dagger_j a_l a_k
\ee
Here the single-particle Hamiltonian $H_{sp}$ is taken with the
well-known  Woods-Saxon form,
\be
\label{vso}
H_{sp} = {p^2\over 2 m} + V_{WS}f(r) + \vec \ell \cdot \vec s\, V_{so}{1\over r}
{d f(r)\over dr}
\ee
with $f(r)=1/(1+\exp((r-R)/a)$ and $R$ the radius of the potential.  We
use parameter values $V_{WS}=-50$ MeV, $a= 0.67$ fm which are close to
standard values in the literature~\cite[Eq.~2.181]{BM}.  We do not expect that the 
competition between the different kinds of pairing will depend sensitively
on $V_{WS}$ and $a$.

   The spin-orbit strength parameter $V_{so}$ will be important in the
competition, and its determination requires some discussion.  Unfortunately,
first-principles studies of the nuclear Hamiltonian are not at a point where
quantitatively reliable spin-orbit fields can be calculated~\cite{tensor}.   
However, given the Woods-Saxon form in Eq. (\ref{vso}), one can constrain
the parameter quite well from spectroscopic data.
A good source is the energies of single-particle excitations in nuclei 
adjacent to closed shells.
For the present study we fit the experimental spin-orbit splitting of the
$f_{7/2}$ and $f_{5/2}$ in the nucleus$^{41}$Ca~\cite{uo94}.  The measured
value is 7.0 MeV, and the   
fitted spin-orbit strength is $V_{so}=33$ MeV-fm$^2$.  We estimate an 
uncertainty of not more than 10\% in this value, based on comparison with
other ways to estimate it.

We only
treat nuclei with equal numbers of neutrons and protons in this work, 
enforcing the condition by taking equal chemical potential in the HFB equations.  
The value of the
chemical potential is fixed by demanding that the density
goes to the standard value  $\rho = 0.16$ fm$^{-3}$ in the nuclear-matter
limit.  With the chosen single-particle Hamiltonian, the corresponding
chemical potential is $\lambda = -13 $ MeV.   
Note that our Hamiltonian neglects the Coulomb interaction.   This allows
us to approach the nuclear matter limit but the calculated nuclei
extend to the unphysical region beyond the proton dripline.

\subsection{Interaction}
The two-particle matrix elements are determined by a contact interaction of 
the general form
\be
\label{v}
\lb ij | v | kl \rb = \sum^6_\alpha v_\alpha \lb ij|\delta^{(3)}(\vec r-\vec r')P_{L=0}
P_\alpha | kl\rb.
\ee
Here $P_{L=0}$ is a projection operator on total orbital angular momentum
zero. The $P_\alpha$ is a projection operator on the six spin-isospin channels 
$(S_{S_z},T_{T_z})= (0_0,1_1),(0_0,1_0),(0_0,1_{-1}),(1_0,0_0),(1_1,0_0),
(1_{-1},0_0)$.
The interaction strengths $v_\alpha$ are of course independent of the
$z$-projections of spin and isospin, and we shall call the two independent
strengths $v_s$ and 
$v_t$ for spin-singlet and spin-triplet, respectively.  However, we will 
need to keep explicit the six amplitudes with different $z$-projections
when constructing the HFB condensate.  It should also be
noted that the contact interaction has to be regulated in some
way to produce a finite condensate.  This will be implemented  
by including only orbitals whose orbital single-particle energies
$\varepsilon_\ell$ are within a range $E_B/2$ of the Fermi energy,
$\lambda-E_B/2 < \varepsilon_\ell < \lambda+E_B/2$.

Our choices for the parameters will be guided by the relevant 
shell model matrix elements of phenomenological shell-model 
Hamiltonians.  In this way we hope to avoid some of the ambiguity 
associated with the form of the pairing interaction (see 
concluding section). 
Specifically, we take the USDB Hamiltonian fitted to $sd$-shell 
nuclei \cite{br06} and the GX1A Hamiltonian fitted to 
$fp$ shell nuclei \cite{ho05}.  We consider all
the shell-model matrix elements with total spin and
isospin couplings $(J,T)=(0,1),(1,0)$, and make a least-squares
fits to each set using the contact interaction and harmonic oscillator
orbitals.  In the fit, we give equal weight to all matrix elements in 
the $m$-scheme representation.  The results for the ratio of interactions 
strengths $v_t/v_s$, shown in Table I, are in the range 1.6-1.7.
Another fitted Hamiltonian to $fp$-shell
spectra \cite{po98} gives a very similar ratio (1.56).  
To get the absolute values of the
strengths, we also need to specify the oscillator parameter controlling
the size of the orbitals.  This is taken as $\hbar \omega = 14$ MeV
and 11 MeV for the $sd$-shell fit and the $fp$-shell fit respectively.
The deduced strength parameters $v_s,v_t$ are also given in the 
table.  One sees that the interactions derived from USDB and GX1A are in
good accord with each other. Unaccountably, the interaction from ref. \cite{po98} has much
lower strengths.
For application   
to the HFB theory outside given shell-model spaces, it is still       
necessary to choose a suitable space cutoff parameter $E_B$.  An     
argument can be made identifying $E_B$ with the harmonic oscillator
shell energy $\hbar\omega$, which would put it in the range of 
$11-14$ MeV.for the strengths in Table I.  
For our study here, we have somewhat arbitrarily taken the strength of 
the spin-singlet and spin-triplet interactions to be 
$v_s=300$ MeV-fm$^3$ and $v_t=450$ MeV-fm$^3$, respectively, with
a cutoff of $E_B=10$ MeV.  The ratio is 1.5, giving the spin-triplet
somewhat less of an enhancement than the fits suggests.  The
calculations will then be on the conservative side in testing
for a possible phase transition.

\begin{table}[h!]
\caption{Strengths of triplet and singlet interactions
from shell-model fits and their ratios.  See text for details.
}
\begin{tabular}{|c|cc|c|}
\colrule
Source &    $v_s$   & $v_t$ &  Ratio \\
   &  MeV-fm$^{3}$ & MeV-fm$^{3}$ & \\
\colrule
$sd$-shell \cite{br06}  & 280. & 465. & 1.65 \\
$fp$-shell   \cite{ho05}   & 291.   & 475.  & 1.63 \\
\colrule
\end{tabular}
\end{table}
\subsection{HFB}
The physical quantity to be calculated is the energy of the system in
a state constructed by making a Bogoliubov transformation on the noninteracting 
ground state.  The expression for the energy is 
\be
E = \sum_{ij} \rho_{ij}<i|H_{sp}|j> + \sum_{i>j} \sum_{i'>j'} \kappa_{ij}
\lb ij | v | i'j'\rb \kappa_{i',j'}
\ee
where $\rho_{ij}$ and $\kappa_{ij}$ are the ordinary and 
anomalous densities associated with the transformation.
There is also a $\rho^2$ term in the interaction energy which we ignore
along with all other Hartree-Fock contributions.
Another convenient quantity for studying the pairing phases is the
condensation energy, also
called the correlation energy $E_{corr}$. This is defined as 
\be
\label{ecorr}
E_{corr} = E_0 -E.
\ee 
where $E_0$ is the energy of the ground state in the absence of a pairing condensate,
The HFB ground state will maximize the correlation energy.
   By the Bloch-Messiah theorem \cite[p. 611-612]{RS}, the solution of the HFB equations
could also be found by the BCS minimization with the variational
wave function $|BCS\rb = \prod_\alpha (u_\alpha +v_\alpha a^\dagger_\alpha
a^\dagger_{\bar \alpha}|\rb$, provided one knows the orbitals $\alpha$
and the pair correspondence $\alpha \rightleftharpoons \bar \alpha$.
For ordinary spin-zero pairing in nuclei with even
numbers of nucleons, we know that the pairs
are the partner orbitals under the time-reversal operator.  If the
orbital basis does not have more than one orbital of given orbital angular
momentum $\ell$, the partners are related by the time-reversal operator,
$ |\bar\alpha\rb = | \ell s j\, {-j_z}\rb$ for
$|\alpha\rb =  | \ell s j j_z\rb$.  Thus, for narrow energy-band
windows $E_B$ and not too large nuclei, the BCS theory is adequate.
This will provide one check on the more complicated HFB
calculations.

   For spin-triplet pairing in the presence of a spin-orbit field, the
HFB treatment is unavoidable because the optimum orbital basis $|\alpha\rb$ 
depends on the relative strengths of the pairing and 
spin-orbit fields.  However, there is one case where the BCS treatment
is valid.  That is for a single $j$-shell and pairing in the 
$S_z=0$ channel.  In that case the pairing field demands that
$j_z(\alpha)= - j_z(\bar \alpha$).  This is uniquely satisfied
by the same partnering as in the $S=0$ pairing channel.  We
shall also use this case for testing the full HFB code.

Another important physical quantity is the quasiparticle energy $E_q$, 
defined as the eigenvalue of the HFB matrix equation, Eq. (\ref{HFB}).  It has the
interpretation as the removal or addition energy for a particle in 
system with an even number of nucleons.

\subsection{The $SU(4)$ limit} 
Before presenting calculations with
physical values of the Hamiltonian parameters, we 
examine the theory in the limit of $SU(4)$ symmetry. The symmetry requires that the
two pairing strengths be equal and that the potential be purely central.  
The HFB ground state of this Hamiltonian is highly degenerate, with many 
distinct ways
to form the pairs.  For example, one solution of the HFB equations is the
state having both NN and PP pairing, with independent condensates for
both.  But we can equally well form a condensate pairing
up-spin neutrons with up-spin protons and another condensate pairing the
down-spin nucleons with each other.  Both states have the same energy in the
$SU(4)$ limit. This will be shown explicitly the example presented in
Sect. IV.  More generally, there is a continuous group of transformations that
leaves the condensation energy invariant.  The degeneracy of the HFB ground
state can only be broken by treatments of the wave function beyond the
HFB approximation, e.g. \cite{en97}.
\section{Calculational details} 
We set up HFB equations in a basis of states
$|i\rb$ constructed from the eigenvectors of the central Hamiltonian $p^2/2m
+ V_{WS}F(r)$. We assume spherical symmetry and represent the orbital wave
functions by their quantum numbers $i=(n,\ell,\ell_z,s_z)$ and radial wave
functions $\phi_i(r)$ on a uniform radial mesh.  The basis is truncated by keeping
only states whose single-particle energies $\varepsilon_\ell$ (without spin-orbit)
are within a range $E_B /2$ of the Fermi energy, $ | \varepsilon_\ell - e_F | <
E_B/2$.  The spin-orbit interaction is treated by including its orbital
interaction matrix elements in the HFB Hamiltonian. 

To efficiently calculate the pairing field  elements $\Delta_{ij}$ 
in the HFB matrix, we save and store the
anomalous densities $\kappa_\alpha(r)$ on a radial grid.
The densities  $\rho_{ij}$ and $\kappa_{ij}$ are calculated in the
orbital representation \cite[p. 251, eq. (7.23)]{RS} from the selected columns of the transformation matrix 
that diagonalizes the HFB Hamiltonian.
Under the assumption of spherical spatial symmetry, the  HFB 
matrix is block-diagonal with respect to the orbital quantum number $\ell$;
the different $\ell$-blocks are calculated separately in our codes.
If one restricts the anomalous density to channels having $S_z=0$, 
the matrix can further be decomposed
into blocks of fixed $|j_z|=|\ell_z+s_z|$.   Further details are 
given in Appendix A.

\section{Results}
\subsection{An example: $^{48}$Cr}
Before applying the theory to very large nuclei, we investigated its
performance in an experimentally well-studied region, namely the mass
region corresponding to an open $f_{7/2}$ shell.  The middle nucleus
filling the $f_{7/2}$ shell is 
$^{48}$Cr, with 4 neutrons and 4 protons outside the completely filled
shells of $1s,1p,2s$, and $1d$ orbitals.  To model this nucleus, we
take the radius parameter $R=4.62$ fm corresponding to the phenomenological
The single-particle Hamiltonian
gives a spectrum that corresponds very well with the shell assignments
with the $f_{7/2}$ shell at the Fermi energy.  The HFB equations are
solved adjusting the chemical potential to get the correct particle
number, $A=48$.
Some results for the HFB theory are presented in Table II, showing how the gaps
and correlation energies depend on the Hamiltonian properties.
We first demonstrate the degeneracy of HFB solutions when the Hamiltonian
has $SU(4)$ symmetry in the spin and isospin degrees of freedom.  The
required Hamiltonian has spin-field field set to zero and equal pairing
strengths $v_s$ and $v_t$.  The calculated properties of the condensate
are shown in the first four rows of Table II.
One sees that the calculated quasiparicle energy is 
independent of the quantum numbers of the condensate and is the same 
for all orbitals.  The correlation energy is also independent of the
choice of condensate, demonstrating the degeneracy of the $SU(4)$ HFB ground 
state.

Going now to a more realistic Hamiltonian, rows 5 and 6 in the Table
show the maximum effect of the spin-orbit field,
taking only the $f_{7/2}$ shell for the single-particle space.  
Row 5  shows the results for an ordinary spin-singlet
condensate, pairing neutrons with neutrons and protons with protons.  The 
quasiparticle energy $E_q$ is the same for all 8 orbitals in the shell.  The
value, $E_q \approx 1$ MeV, is in fact 
close to the average experimental odd-even mass difference.  
Row 7 shows the results for spin-triplet pairing.
The $S=1$ condensates define a direction in space and the gaps are no longer
independent of the $j_z$ quantum number of the orbitals.  These gaps are
plotted as a function of $j_z$ in Fig. 1.  One sees that the
gap approaches zero
as $j_z$ becomes small.  This behavior would be called gapless 
superfluidity in a large system.  The correlation energies of spin-singlet
and spin-triplet pairing can be compared in the last column of the 
Table. One sees that it smaller for spin-triplet than for spin-singlet.
Thus, the ground state should exhibit ordinary pairing, as expected.

We next augment the single-particle space by adding the $f_{5/2}$ shell,
 including the physical spin-orbit splitting.  The results for four 
choices of condensate are
presented.  First of all, one sees that the correlation
energies are larger than in the pure $f_{7/2}$ Hamiltonian.  They could
not be smaller
because the HFB theory is variational: enlargening the space can only
lower the energy.  The first two rows here
show the results for spin-singlet pairing, differing on the isospin
coupling.  The energy are exactly the same due to the
isospin invariance of the Hamiltonian.  A similar degeneracy is
found for the spin-triplet
pairing; here the condensate energy does not depend on $S_z$.  
We shall exploit this invariance later by limiting our trial condensates
to be either $(0,1_0)$ or $(1_0,0)$.  Then $j_z$ is conserved and the 
HFB matrix can be diagonalized in small blocks (See Appendix A).
The final entry in the Table is for the Hamiltonian in the full
$fp$-shell.  This is in fact the truncation that result from
the $E_B=10$ MeV cutoff around the Fermi energy.  We see that the
correlation energies are larger, as expected.  The spin-singlet
pairing is still the stronger one, so the Hamiltonian with the 
$E_B$ passes the test of agreeing with known phenomenology.
\begin{table}[h!]
\caption{Pairing gaps and correlation energies for $^{48}$Cr in the HFB theory.
$D$ is the dimension of the $\Delta$ or $\kappa$ matrix; the dimension
of the HFB matrix is twice that.  The spin-singlet interaction strength
is $v_s=300$ MeV-fm$^3$.  The spin-triplet interaction strength is 
$v_t=450$\Mf~except for the rows with the $SU(4)$ results; there 
$v_t=300$\Mf. The condensates are labeled by the spin-isospin 
quantum numbers, dropping the $z$-quantum number when $S=0$ or $T=0$.
The column marked $E_q$ gives the value or the range of values of the
quasiparticle energy of orbitals at the Fermi energy. 
The correlation energy $E_{corr}$ in
the last column is defined in eq. (\ref{ecorr}).
}
\begin{tabular}{|c|c|c|cc|}
\colrule
Space & D   & condensate  &  $E_q$ & $E_{corr}$\\
      & &$(S_{S_z},T_{T_z})$ & (MeV) & (MeV) \\
\colrule
$f$-shell & 28  & $(0,1_1),(0,1_{-1})$& 1.80 &  10.27  \\
$SU(4)$  &  & $(0,1_0)$& 1.80 &  10.27  \\
  & &$(1_1,0),(1_{-1},0)$& 1.80 &  10.27  \\
 & &$(1_0,0)$& 1.80 &  10.27  \\
\colrule
$f_{7/2}$ & 16  & $(0,1_1),(0,1_{-1})$ & 1.03 & 4.11 \\
 &  & $(0,1_0)$ & 1.03 & 4.11 \\
  &  & $(1_0,0)$ & 0.12-0.87 & 2.01\\
\colrule
$f_{7/2},f_{5/2}$ & 28 & $(0,1_1),(0,1_{-1})$ & 1.16 & 4.64 \\
 & &$(0,1_0)$ & 1.16 & 4.64 \\
 & &$(1_1,0),(1_{-1},0) $ & 0.20-1.34 & 2.98  \\
 & &$(1_0,0)$ & 0.20-1.34 & 2.98 \\
\colrule
full $fp$ & $28+12$ &$(0,1_1),(0,1_{-1})$  & 1.23 & 4.92 \\
 & &$(0,1_0)$  & 1.23 & 4.92\\
 &  &$(1_1,0),(1_{-1},0)$  & 0.26-1.58 & 3.45 \\
 & &$(1_0,0)$   & 0.26-1.58 & 3.45 \\
\colrule
\end{tabular}
\end{table}

\begin{figure}
\includegraphics [width = 9 cm]{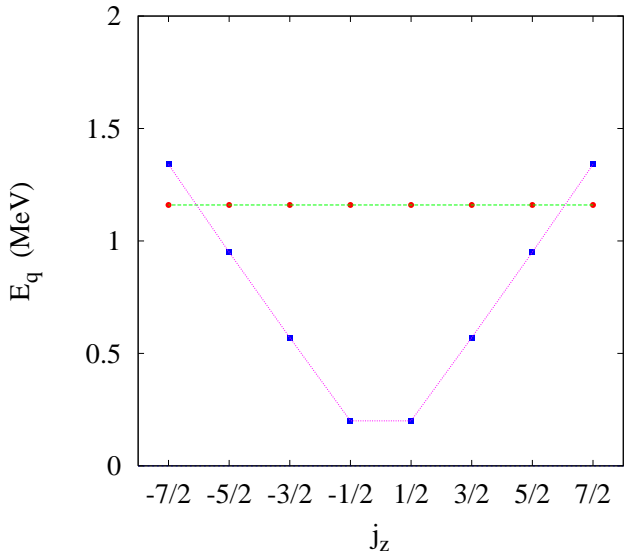}
\caption{\label{gaps} Quasiparticle energies in $^{48}$Cr for the
$f$-shell space.  Red circles: spin-singlet; blue squares: spin-triplet
with condensate in the $S_z=0$ channel.  Lines are drawn to guide the eye.
}

\end{figure}
\subsection{Systematics}
In this section, we compare correlation energies for the triplet and
singlet pairing channels as function
of nuclear size.    We showed in the previous section that the correlation
energy does not depend on the $z-$components of the spin or isospin
of the condensate, so we may limit our exploration of possible
condensates to the $S_z=0$ channels. As mentioned above, the HFB matrix reduces to blocks of fixed $|j_z|$ for these
channels. We find the self-consistent solutions of the HFB equations iterating
from a starting point in which there is a finite condensate in some
channel.   

For the range of nuclear sizes that
we consider, the pairing gap is comparable to or smaller than the energy
spacing of the shell orbitals.  Under these conditions the strength of the pairing
condensate will be quite sensitive to the Fermi level.
Since we are interested here in nuclei that have well-formed pairing
condensates, we shall only calculate systems where the chemical
potential ($\lambda= -13$ MeV) coincides with the orbital energy of
some $j$-shell.  This gives about 50 cases for nuclei with
radii in the range of 4.0-12.0 fm.  These correspond to mass numbers
in the range $A=25-1000$.  A table of the results for
the two correlation energies and their ratio is given in 
Appendix B.  For the lighter nuclei, correlation energies are
of the order of several MeV, except for several nuclei with
$j=1/2$ shells at the Fermi level, for which the spin-singlet correlation
energy can be less than 1 MeV.  In the heavy nuclei, the correlation
energy can be several tens of MeV.  

Fig. 2 shows a plot of the ratio of spin-triplet to spin-singlet
correlation energies as a function of the mass $A$.
The results show that there is a    
trend to favor spin-triplet pairing for large nuclei, as was argued
in the Introduction.  However, the ratio is by no means monotonic  
as a function of $R$.  Spin-singlet is favored for $R<6.5$ fm and  
spin-triplet for $R>9$ fm, but nuclei in between could have 
either ground state.  
The lightest nuclei predicted to have spin-triplet condensates
are indicated by their mass and element designation. They are 
$^{30}$P, $^{76}$Sr, and $^{136}$Er.
The two lighter ones are in the physical region,
and the spectroscopic properties of one of them,  $^{30}$P, are well-known.
The calculated correlation energy for this nucleus is 1.4 MeV, which is
not large enough to be considered a strong condensate.  Nevertheless,
the observed ground state spin and parity  ($J^\pi =1^+$) agrees with the 
quantum numbers of the spin-triplet condensate.  The middle nucleus,
$^{76}$Sr has a somewhat smaller condensation energy, $E_{cor}\approx 1$
MeV, so any effects of the pairing would be quite weaker.  The
nuclei in this mass region are accessible with current-generation
accelerators and their experimental spectroscopic properties are under active 
study.  The third nucleus, $^{136}$Er, just beyond the
physical region, 
is the smallest nucleus having
both strong pairing ($E_c\sim 13$ MeV) and a spin-triplet condensate.  There are two
nearly degenerate shells
at the Fermi energy, the $2s_{1/2}$ and $1d_{3/2}$ shells. They both have 
radial nodes and weak spin-orbit splitting from their $j$-shell partners.

\begin{figure}
\includegraphics [width = 9cm]{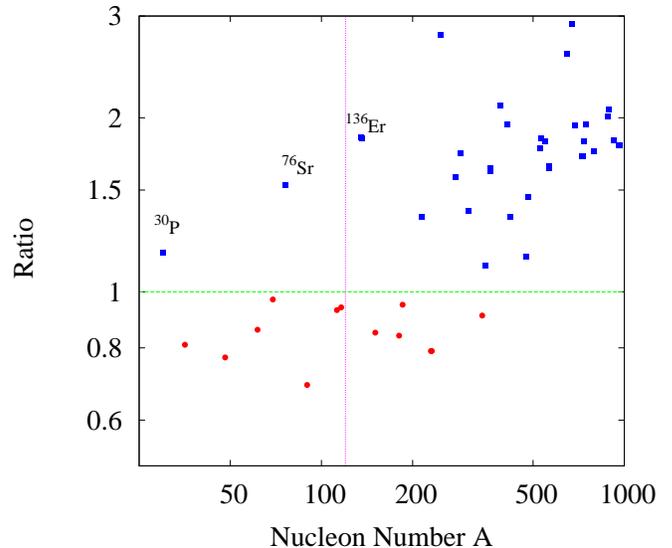}
\caption{\label{ratio} Ratio of spin-triplet to spin-singlet correlation
energies as a function of mass number $A$.  Nuclei with spin-singlet and
spin-triplet condensates are shown as red circles and blue squares,
respectively.  The vertical line at $A\approx 120$ shows the dividing
line between nuclei that are bound (left) and nuclei that unstable with
respect to proton emission, according to the mass table of ref. \cite{mn}. 
}
\end{figure}

A table of the correlation energies is given in Appendix B.  The table
also shows the angular momentum quantum numbers at the Fermi energy
for each case calculated.

\section{Conclusion}
This study offers a resolution of the conundrum, why is it that 
ordinary nuclei only exhibit spin-singlet pairing when nuclear matter 
calculations find that spin-triplet
pairing is stronger?  The answer, that ordinary nuclei are too small
to be out of the influence of the surface spin-orbit field,
is seen by calculating the pairing energies over a large range of
sizes.
For our Hamiltonian, the spin-triplet pairing dominates for all calculated
nuclei whose radii are larger than $\sim9$ fm, and for a number of
lighter nuclei having low-$l$ shells at the Fermi energy.  The candidate
for the smallest nuclei with a well-developed condensate, $A\sim 130-140$,
is tantalizingly close to region of physical nuclei defined by the
single-nucleon driplines.  

However, due to the simplifications made to the Hamiltonian and to other
uncertainties we cannot be quantitative about the transition point.  
In particular, important physics may be lost in 
ignoring the momentum- and density-dependence of the pairing interaction.
In the theory of nuclear matter the spin-singlet pairing 
derived from Brueckner theory is very weak at nuclear matter density 
\cite[See Fig.~8]{ba90}\cite{ga01}; see also the contradicting claim in ref.
\cite{el96}. The pair gap becomes larger at subnuclear densities,  suggesting that
surface effects are crucial for that channel.  In many studies of nuclear
structure, the authors assume that there is some surface enhancement of 
the pairing, see e.g. \cite{te95,do01,sa05}.  Beyond the effects that
are treated in the Brueckner theory,  there is likely to be a significant  
induced interaction associated with  collective
surface vibrations \cite{Broglia}.  If we included either mechanisms
enhancing the surface region, the pairing energy would decrease as
the nucleus becomes larger.  This of course could effect the crossover
from singlet to triplet condensates.   Also one needs to take into
account renormalization effects associated with the truncation of 
the orbital space.  One finds that the renormalized interaction is
actually suppressed in the surface region \cite{yu03}.  Clearly,
there is much left to do to improve the interaction.

It would also be interesting to explore the effects of nuclear deformation.
Breaking the  spherical symmetry would effectively reduce the single-particle level 
density at mid-shell filling, weakening the pairing condensates.  How
this affects the competition between the two kinds of pairing is
not known.  

There are a number of other questions that could be addressed
within the context of the simplified Hamiltonian that we have employed.
Can two kinds of condensates coexist in the same system?    At first
sight this seems
unlikely because the bulk pair correlation length
is large compare to the sizes of the nuclei we consider.  However,
it has been claimed that the Cooper pairs have a small size in the
nuclear surface\cite{pi07}.  This might allow the existence of a mixed phase with
a strong spin-triplet condensate in the interior and a spin-singlet
condensate on the surface.  Another interesting question is the fragility
of spin-triplet pairing is when there is a neutron excess.  For our
midmass triplet-pairing candidate nucleus, $N=Z\sim 65$, the condensate
seems quite robust with a calculated 
correlation energy 6 MeV larger than the singlet energy. It might be
that the 
system could support a few excess neutrons and still preserve the triplet
condensate.

We have not addressed the question of how the triplet pairing condensate
could be observed, beyond noting that the character of the condensate
controls quantum numbers
of the ground state in an odd-odd system.
In fact several $J=1,T=0$ ground states are sprinkled among the 12 odd-odd 
$N=Z$ 
nuclei known experimentally.  The $\beta$-decay spectrum may also
be affected by the pairing\cite{en97}.  Finally, two-particle transfer
reactions would be affected by the coherence of the pairing fields.

The authors thank M. Horoi for supplying us with the GX1A interaction.
We also thank A.~Bulgac, T. Duguet, B. Spivak and S. Baroni for discussions, and J. Vinson for help in
an early stage of this work.
This work was supported by DOE Grant DE-FG02-00ER41132.

\section*{Appendix A}
  The Hartree-Fock-Bogoliubov equations have
the general structure 
\be
\label{HFB}
\left[\matrix{ H_{sp} - \lambda \bm{1} & \Delta \cr
 -\Delta^* & -H_{sp} + \lambda \bm{1} }\right] 
\left( \matrix{u_\alpha \cr v_\alpha} \right)
= E_\alpha \left( \matrix{u_\alpha \cr v_\alpha} \right).
\ee
Here $H_{sp}$ and $\Delta$ are matrices in some basis of orbital wave
functions and $u_\alpha,v_\alpha$ are column vectors.
The ordinary and anomalous density matrix elements are given by
\be
\rho_{ij} = \sum_\alpha v^*_{\alpha,i}v_{\alpha,j}
\ee
$$
\kappa_{ij} = \sum_\alpha u^*_{\alpha,i}v_{\alpha,j}
$$
where the sums are restricted to eigenvectors satisfying
$E_\alpha > 0$.  We construct the HFB matrix assuming that
the nucleus is spherical with spherically symmetric fields
except for the spin variables, using as a basis of the orbital
wave functions the eigenfunctions of the central potential which 
may be labeled by the quantum numbers $(n,\ell,\ell_z,s_z,t_z)$. 

To efficiently calculate the pairing field  elements $\Delta_{ij}$ 
in the HFB matrix, we save and store the
anomalous densities $\kappa_\alpha(r)$ on a radial grid.  The densities are defined 
\be
\kappa_\alpha(r) ={1\over 4 \pi} \sum_{i>j} \phi_i(r)\phi_j(r) \lb 
\alpha | ij\rb \kappa_{ij}
\ee
where 
\be
\lb i j | \alpha\rb = \sqrt{2} 
\left((1/2\, s_{zi}\, 1/2 \,s_{zj} | S_\alpha S_{z\alpha}\right)\times
\ee
$$
\left(1/2\, t_z(i)\, 1/2\, t_z(j) | T(\alpha) T_z(\alpha)\right)
 (-)^{\ell_i-\ell_{zi}}
\delta_{\ell_i,\ell_j}\delta_{\ell_{zi},\ell_{zj}}.
$$
and the radial wave functions $\phi_i$ are normalized as
$\int_0^\infty r^2 dr |\phi_i(r)|^2 = 1$.
The elements of the
HFB $\Delta$ matrix are calculated as
\be
\label{delta}
\Delta_{ij} = \sum_\alpha \lb ij | \alpha \rb v_\alpha \int_0^\infty
r^2 d r  \phi_i(r) \phi_j(r) \kappa_\alpha(r).
\ee 
where $v_\alpha$ is the strength of the contact interaction defined
in eq. (3).

Due to the
symmetry, 
the condensation energies can be calculated under the 
further assumption that the total azimuthal angular momentum 
$|j_z|=|el_z+s_z|$ is conserved.  The HFB matrix then decomposes into $2\times 2$
or $4 \times 4$ dimensional blocks with respect to spin and
orbital angular momentum $s_z$ and $\ell-z$.
A typical subblock for a fixed isospin of the pairing field has the
form
\be
\left[\matrix{\bm{h_{\ell,j_z}} - \lambda\bm{1} & 0 & 0& \bm{\Delta_\ell} \cr
              0 & \bm{h_{\ell, -j_z}} - \lambda\bm{1} & -\bm{\Delta_\ell} & 0 \cr
              0 &-\bm{\Delta_\ell} &  -\bm{h_{\ell,jz}} + \lambda\bm{1} & 0\cr
              \bm{\Delta_\ell} & 0  & 0 & -\bm{h_{\ell,-j_z}} + \lambda\bm{1}\cr}\right]
\ee
Assuming that there is only one radial state in the $E_B$ window,
the $\bm{h}$ and $\bm\Delta$ matrices have 
dimension 1 or 2 depending
on whether $|j_z|=\ell+1/2$ or not.  In the latter case, the 
$\bm{h_{\ell,j_z}}$ matrix is  given by
\be
\bm{h_{\ell,j_z}} =\left[\matrix{ \varepsilon_\ell+(j_z-1/2)w_\ell/2 &
   \sqrt{(\ell+1/2)^2-j_z^2}w_\ell/2    \cr
             \sqrt{(\ell+1/2)^2-j_z^2}w_\ell/2&  \varepsilon_\ell-(j_z+1/2)m 
w_\ell/2 \cr}\right]
 \ee
where $\varepsilon_\ell$ is the orbital energy calculated without
the spin-orbit field and $w_\ell$ is the radial matrix element of
the spin-orbit field.
The  matrix $\bm\Delta_\ell$ is given by 
\be
{\bm\Delta_\ell} = \left[\matrix{ 0 &  d_\ell  \cr
             \pm d_\ell  & 0 \cr}\right].
\ee 
Here the upper and lower signs apply  to the $(ST)=(01)$ and $(10)$ channels,
respectively.  The pairing field $d_\ell$ is identical to $\Delta_{ij}$
in eq. (\ref{delta}) with eg. $|i\rb=|n\ell \ell_z s_z t_z\rb$ and
$|j\rb=|n\ell\,{-\ell_z} \,{-s_z}\, {-t_z}\rb$.
Note that $d_\ell$ is the same for all the HFB subblocks of given
$\ell$. The $2\times2$ HBF matrix in the $|j_z|=\ell+1/2$
case can be obtained by eliminating the orbitals with unphysical
$m$ values from the basis.

To solve the HFB equations, we start with a trial field for the
anomalous density $\kappa_\alpha(r)$, typically a constant value 
independent of $r$ in some channel $\alpha$ and zero in the
other channels.  The solution of the HFB equation with the
trial field is used to calculate a new field, and the process
is iterated to convergence.  

A computer code to solve the HFB equations based on these formulas 
is available by E-mail from the authors, {\tt bertsch@u.washington.edu}.

\section*{Appendix B}
Columns in the table refer to:\\
$R$, the radius parameter of the Woods-Saxon potential eq. (2)\\
$A$, the nuclear mass number\\
$E_c(s)$, the spin-singlet condensation energy in MeV\\
$E_c(t)$, the spin-triplet condensation energy\\
$R_{ts}=E_c(t)/E_c(s)$, their ratio\\
$\ell,j$, orbital and total angular momentum of the single-particle orbital at
the Fermi energy.

{\tt
~R~~~~~~~A~~~~Ec(s)~Ec(t)~~R\_ts~~~l~~~j\\
-----------------------------------------\\
4.14~~~~29.9~~1.3~~~1.5~~~1.17~~~~0~~~1/2\\
4.28~~~~35.4~~2.4~~~2.0~~~0.81~~~~2~~~3/2\\
4.77~~~~48.1~~4.8~~~3.7~~~0.77~~~~3~~~7/2\\
5.20~~~~61.5~~2.8~~~2.4~~~0.86~~~~1~~~3/2\\
5.30~~~~69.1~~4.4~~~4.3~~~0.97~~~~3~~~5/2\\
5.40~~~~76.2~~0.9~~~1.3~~~1.53~~~~1~~~1/2\\
5.70~~~~89.7~~5.1~~~3.5~~~0.69~~~~4~~~9/2\\
6.24~~~112.5~~5.9~~~5.5~~~0.93~~~~2~~~5/2\\
6.27~~~116.2~~6.6~~~6.2~~~0.94~~~~4~~~7/2\\
6.50~~~135.1~~7.0~~12.9~~~1.85~~~~0~~~1/2\\
6.51~~~135.7~~7.0~~12.9~~~1.84~~~~2~~~3/2\\
6.62~~~150.6~~6.5~~~5.6~~~0.85~~~~5~~~11/2\\
7.21~~~180.5~~6.6~~~5.6~~~0.84~~~~5~~~9/2\\
7.24~~~185.3~~5.8~~~5.6~~~0.95~~~~3~~~7/2\\
7.51~~~214.8~13.5~~18.3~~~1.35~~~~6~~~13/2\\
7.57~~~229.8~~7.4~~~5.8~~~0.79~~~~3~~~5/2\\
7.58~~~231.5~~7.3~~~5.8~~~0.79~~~~1~~~3/2\\
7.72~~~247.3~~0.8~~~2.3~~~2.78~~~~1~~~1/2\\
8.13~~~276.8~~9.5~~15.0~~~1.58~~~~6~~~11/2\\
8.23~~~287.0~~9.8~~17.1~~~1.74~~~~4~~~9/2\\
8.39~~~305.1~11.8~~16.3~~~1.38~~~~7~~~15/2\\
8.60~~~339.5~~3.9~~~3.6~~~0.91~~~~4~~~7/2\\
8.64~~~346.6~~3.3~~~3.6~~~1.11~~~~2~~~5/2\\
8.84~~~360.9~~1.9~~~3.1~~~1.62~~~~2~~~3/2\\
8.85~~~361.5~~1.8~~~3.0~~~1.64~~~~0~~~1/2\\
9.03~~~390.8~12.3~~25.7~~~2.10~~~~7~~~13/2\\
9.19~~~410.7~15.7~~30.5~~~1.95~~~~5~~~11/2\\
9.26~~~420.9~14.0~~18.8~~~1.35~~~~8~~~17/2\\
9.60~~~474.1~~4.2~~~4.9~~~1.15~~~~5~~~9/2\\
9.68~~~481.9~~3.7~~~5.4~~~1.46~~~~3~~~7/2\\
9.92~~~528.0~17.9~~31.6~~~1.77~~~~8~~~15/2\\
9.95~~~532.8~16.7~~30.7~~~1.84~~~~1~~~3/2\\
10.05~~~549.8~19.7~~35.8~~~1.82~~~~1~~~1/2\\
10.13~~~562.8~23.2~~38.0~~~1.64~~~~9~~~19/2\\
10.14~~~564.5~22.9~~37.8~~~1.65~~~~6~~~13/2\\
10.57~~~647.2~~7.2~~18.5~~~2.58~~~~6~~~11/2\\
10.69~~~669.4~11.4~~33.1~~~2.91~~~~4~~~9/2\\
10.79~~~688.8~16.5~~32.1~~~1.94~~~~9~~~17/2\\
10.97~~~726.2~27.5~~47.3~~~1.72~~~~4~~~7/2\\
10.98~~~728.2~27.5~~47.2~~~1.72~~~10~~~21/2\\
10.99~~~730.3~27.3~~47.0~~~1.72~~~~4~~~7/2\\
11.03~~~738.6~24.8~~45.1~~~1.82~~~~2~~~5/2\\
11.07~~~747.1~21.1~~41.1~~~1.95~~~~7~~~15/2\\
11.18~~~791.4~10.2~~18.0~~~1.75~~~~2~~~3/2\\
11.53~~~854.6~10.8~~36.9~~~3.41~~~~7~~~13/2\\
11.66~~~882.0~22.6~~45.3~~~2.01~~~10~~~19/2\\
11.69~~~888.4~21.8~~45.0~~~2.07~~~~5~~~11/2\\
11.83~~~920.2~19.1~~34.9~~~1.83~~~11~~~23/2\\
11.99~~~960.7~18.3~~32.7~~~1.79~~~~8~~~17/2\\
12.00~~~963.2~18.2~~32.6~~~1.79~~~~5~~~9/2\\
}

\begin{thebibliography}{99}
\bibitem{ma00a}A.O.~Macchiavelli, et al., Phys. Rev. C {\bf 61} 041303R
(2000).
\bibitem{ma00b}A.O.~Macchiavelli, et al., Phys. Lett. B {\bf 480} 1
(2000).
\bibitem{po98} A.~Poves and G.~Martinez-Pinedo, Phys. Lett. B{\bf 430}
203 (1998).
\bibitem{baroni} G.F.~Bertsch and S. Baroni, arXiv:0904.2017.
\bibitem{schwenk} G.F.~Bertsch, A.O.~Macchiavelli, and A.~Schwenk,
unpublished; see also S. Baroni, A.O.~Macchiavelli, and A.~Schwenk,
arXiv:0912.0697.
\bibitem{superconductivity} This may be contrasted with the situation
in condensed matter physics:  the electronic spin-orbit field does
not affect the condensation energy.  See e.g. A. Fetter and P. Hohenberg,
in ``Superconductivity", ed. R.D.~Parks, (Marcel Dekker, 1969), vol. II,
p. 882.)  The difference is that the nuclear spin-orbit field affects the
density of states at the Fermi level, but the electronic spin-orbit 
is too weak to do so in metals.
\bibitem{BM} A.~Bohr and B.~Mottelson, ``Nuclear Structure," (Benjamin, 1969), 
Volume I.
\bibitem{tensor}
In particular, the nucleon-nucleon tensor interaction contributes to the
splittings of spin-orbit partners, but with a different radial dependence
than that arising from the nucleon-nucleon spin-orbit interaction.  See
B. A. Brown, T. Duguet, T. Otsuka, D. Abe, T. Suzuki,
Phys. Rev. C74  061303 (2006).
\bibitem{uo94} Y. Uozumi, et al. Phys. Rev. C {\bf50} 263 (1994).
\bibitem{br06} B.A. Brown and W.A. Richter, Phys. Rev. C {\bf 74} 034315
(2006).
\bibitem{ho05} M. Honma, T. Otsuka, B.A. Brown, and T. Mizusaki,
Eur. Phys. J. A {\bf 25} Suppl. 1,499 (2005); Phys. Rev. C {\bf 69}
034335 (2004). 
\bibitem{RS} P. Ring and P. Schuck, ``The nuclear many-body
problem", (Springer, 1980).
\bibitem{mn} P. M\/oller, et al., At. Data Nucl. Data Tables {\bf 9} 185
(1995).
\bibitem{ba90} M.~Baldo, et al., Nucl. Phys. {\bf A515} 409 (1990).
\bibitem{ga01} E.~Garrido, et al., Phys. Rev. C {\bf 63} 037304 (2001).
\bibitem{el96} O. Elgaroy, L. Engvik, M. Hjorth-Jensen, and E. Osnes,
Phys. Rev. Lett. {\bf 77} 1428 (1996); Phys. Rev. C{\bf 57} R1069 (1998).
\bibitem{te95} J. Terasaki, et al., Nucl. Phys. {\bf A593} 1, (1995).
\bibitem{do01} J. Dobaczewski, W. Nazarewicz, and P.-G. Reinhard, 
Nucl. Phys. {\bf A693} 361 (2001).
\bibitem{sa05} N. Sandulescu, P. Schuck, and X. Vin\~as, Phys. Rev.
C {\bf 71} 054303 (2005).
\bibitem{Broglia} A. Pastore, et al., Phys. Rev. C {\bf 78} 024315 (2008).
\bibitem{yu03} Y. Yu and A. Bulgac, Phys. Rev. Lett. {\bf 90} 222501 (2003).
\bibitem{pi07} N.~Pillet, N.~Sandulescu, and P.~Schuck, Phys. Rev. 
C {\bf76} 024310 (2007).
\bibitem{en97} J.~Engel, S.~Pittel, M.~Stoitsov,P.~Vogel, and J. Dukelsky,
Phys. Rev. C {\bf 55} 1781 (1995).
\end{thebibliography}
\end{document}